\documentclass[a4paper,11pt]{article}
\pdfoutput=1 

\usepackage{jinstpub} 

\usepackage{tabularx}

\title{Performance of the ALICE electromagnetic
calorimeters in LHC Runs 1 and 2 and
upgrade projects}

 \author{D. Blau}
 \affiliation{NRC "Kurchatov institute",\\123182 Kurchatov sq. 1, Moscow, Russia}

\emailAdd{dmitry.blau@cern.ch}

\abstract{ALICE (A Large Ion Collider Experiment) at the LHC is designed for studies of nuclear matter at extreme temperatures and energy densities, so called Quark-Gluon Plasma (QGP). Two detectors for measurements of electromagnetic signals, high-granularity photon spectrometer PHOS and large acceptance electromagnetic calorimeter EMCal/DCal, are incorporated into the ALICE apparatus. In order to enhance the data sample of the high-energy part of the spectra, a dedicated triggers on high energetic photons or jets are used. High precision measurements of neutral mesons require fine energy and timing calibrations. Photon signal purity is improved using dedicated photon PID criteria based on cluster shape and anti-track matching with the ALICE central tracking system, as well as a special detector located in front of PHOS called CPV.

PHOS and EMCal participated in LHC Run 1 (2009-2013) and Run 2 (2015-2018) during data taking periods of pp, p-Pb and Pb-Pb collisions. We give an overview of their performance in Runs 1 and 2 in low and high multiplicity environments as well as the upgrade plans for future LHC runs.}


\collaboration[c]{on behalf of ALICE collaboration}

\proceeding{CHEF 2019 - Calorimetry for High Energy Frontier 2019\\
  25-29 November 2019\\
  Fukuoka, Japan}

\begin{document}
\maketitle
\flushbottom

\section{Introduction}
\label{sec:intro}
ALICE (A Large Ion Collider Experiment) at the LHC is designed for studies of nuclear matter at extreme temperatures and energy densities, so called Quark-Gluon Plasma (QGP) \cite{alice}. In order to study electromagnetic signatures of QGP, it incorporates two kinds of electromagnetic calorimeters: highly granulated photon spectrometer PHOS \cite{phos} and large acceptance calorimeter EMCal/DCal \cite{emcal, dcal}. Both are located in the central part of the ALICE detector.

The PHOS spectrometer is an electromagnetic calorimeter based on scintillating PbWO$_{4}$ crystals dedicated for precise measurements of spectra, collective flow and correlations of direct photons, and of neutral mesons in ultra-relativistic nuclear collisions at LHC energies. The choice of active media allowed operation in a high multiplicity environment, the ability to reconstruct neutral pions up to very high transverse momenta and to reach excellent energy and spatial resolutions. In order to increase light yield of crystals and further improve the energy resolution, PHOS is cooled to a constant temperature of -25$^{\circ}$ C. Dedicated L0 and L1 triggers allowed for an increased collected integrated luminosity in data taking. PHOS consists of 3 + 1/2 modules with total of 12544 channels. A Charged Particle Veto (CPV) detector is designed to reject charged particles from the cluster spectrum. First module of CPV was installed to the detector during the Long Shutdown 1 (LS1) between Run 1 and Run 2. Two more modules are being installed and comissioned during current the Long Shutdown 2 (LS2) for future Run 3 (2021-2025).

EMCal is a sampling calorimeter based on lead/scintillator layers used for the measurement of electrons from heavy flavour decays, and the electromagnetic component of jets, spectra and correlations of isolated direct photons and spectra of neutral mesons. During LS1 EMCal acceptance was enlarged with additional modules, DCal, creating possibilities to study di-jets. In total EMCal/DCal has 17644 cells.

Both calorimeters utilize Avalanche PhotoDiodes (APD) S8664-55 produced by Hamamatsu as photodetectors. The readout system is also very similar and is based on digitization and sampling with ADC ALTRO which then pass signal to scalable readout units (SRU) via 20-Mb/s p2p links \cite{readout}. In order to obtain wide dynamic range, two signal amplification modes, high and low gains (HG and LG), are implemented in separate ADCs. The readout time is 172 $\mu$s independent from event size. Comparison of properties of the two calorimeters of ALICE is shown in Table \ref{tab:i}.

\begin{table}[h!]
\centering
\caption{\label{tab:i} Summary of PHOS and EMCal properties.}
\smallskip

  \begin{tabular}{| l | l | l | }
  \hline

  & EMCal/DCal & PHOS \\
\hline
Active element & Sampling 77 layers (1.44 mm Pb, 1.6
mm Sc) & Homogeneous crystals PbWO$_{4}$\\
\hline
Moli\`ere radius & 3.2 cm & 2.0 cm\\
\hline
Photodetector & APD $5\times5$ mm$^2$ & APD $5\times5$ mm$^2$\\
\hline
Depth & 20 $X_0$ & 20 $X_0$\\
\hline
Acceptance & Run 1: EMCal: $|\eta| < 0.7$, $80 < \phi < 180^\circ$ & Run 1: $|\eta| < 0.12$, $260 < \phi < 320^\circ$\\
& Run 2: EMCal: $|\eta| < 0.7$, $80 < \phi < 187^\circ$ & Run 2: $|\eta| < 0.12$, $250 < \phi < 320^\circ$\\
& DCal: $0.22 < |\eta| < 0.7$, $260 < \phi < 320^\circ$ & \\
& $|\eta| < 0.7$, $320 < \phi < 327^\circ$ & \\
\hline
Granularity & Cell $6\times6$ cm$^2$ & Cell $2.2\times2.2$ cm$^2$\\
& $\Delta\phi\Delta\eta$ = $0.0143\cdot0.0143$ rad & $\Delta\phi\Delta\eta$ = $0.0048\cdot0.0048$ rad\\
\hline
Modularity & EMCal: 10+2(1/3) modules & 3+1/2 modules \\
& DCal: 6(2/3) + 2(1/3) modules & \\
& 17664 cells & 12544 cells \\
\hline
Dynamic range & 0-250 GeV & 0-100 GeV\\
\hline
Energy resolution & $\sigma_{E}/E = 4.8\%/E \oplus 11.3\%/\sqrt{E} \oplus 1.7\%$ & $\sigma_{E}/E = 
1.8\%/\sqrt{E} \oplus 3.3\%/E \oplus 1.1\%$ \\
\hline
Distance from IP & 428 cm, 0.7-0.9 $X_0$ & 460 cm, 0.2 $X_0$ \\
\hline
\end{tabular}
\end{table}
\section{Trigger system and its performance}
\label{sec:trigger}
\begin{figure}[ht]
\centering
\includegraphics[width=7cm]{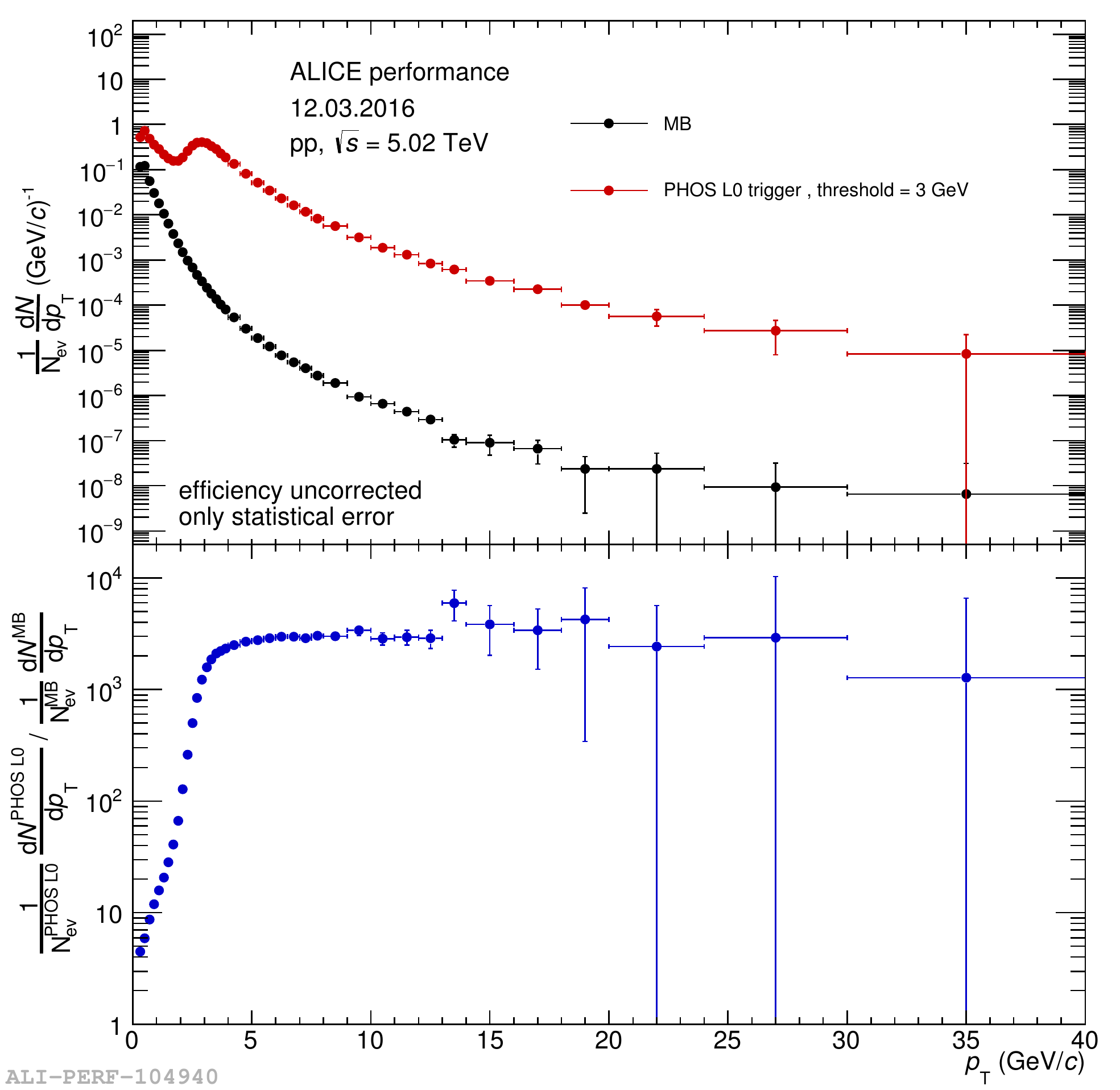}
\includegraphics[width=6cm]{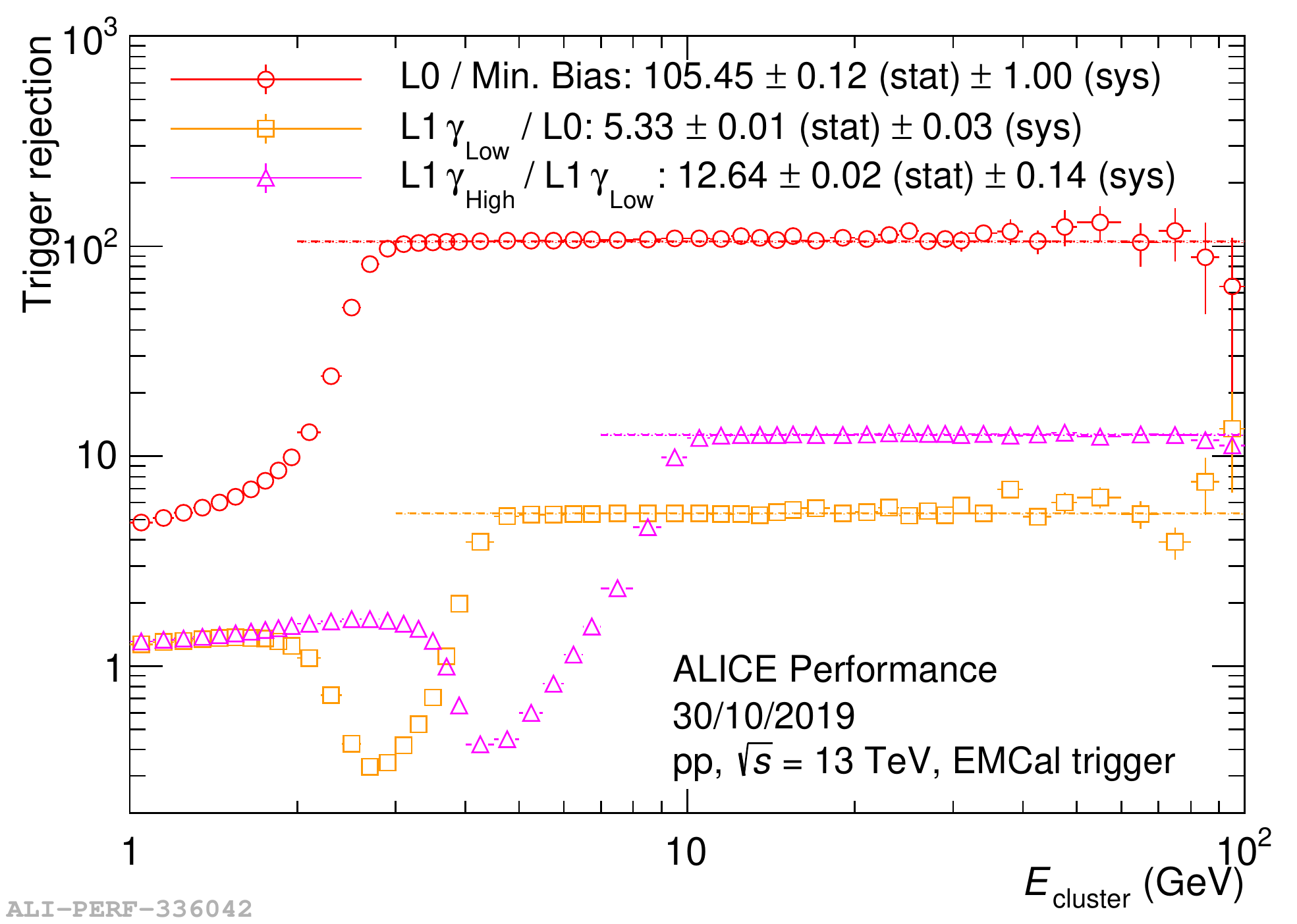}
\caption{Trigger turn-on curves for PHOS and EMCal.}
\label{fig-trigger}       
\end{figure}

ALICE calorimeters provide triggers at levels L0 and L1 in order to increase luminosity of high energy signals in data taking \cite{trigger1,trigger2,trigger3}. Both EMCal and PHOS L0 triggers are fired if the sum of the energy of $4\times4$ cells is larger than a  configurable threshold. In the L0 trigger, the typical thresholds are 2.5-9 GeV. The PHOS L1 trigger provides 3 thresholds which is useful to study Pb-Pb collisions. The EMCal L1 trigger has 2 thresholds on photon signals, and 2 thresholds for jets. Thresholds are adjustable depending on collision rate, required trigger rejection factor, readout time. Typical turn-on curves for L0 and L1 triggers in PHOS and EMCal are shown in Figure \ref{fig-trigger}.
%
\section{Detector calibration}
\label{sec:calibration}
Excellent energy and time resolutions of detectors are crucial for measurements of neutral mesons and direct photons, especially in the low-$p_\mathrm{T}$ range. In order to achieve this, four mutually dependent procedures for energy calibration were performed: relative gain calibration, absolute energy calibration, nonlinearity correction, and time-dependent calibration
correction \cite{calib}. For relative gain calibration in PHOS, pre-calibration based on APD gain equalization and fine energy calibration based on $\pi^0$ peak position were used. Absolute energy scale was checked with electrons $E$/$p$ in PHOS, and with electron beam tests in case of EMCal. 

\begin{figure}[ht]
\centering
\includegraphics[width=5cm]{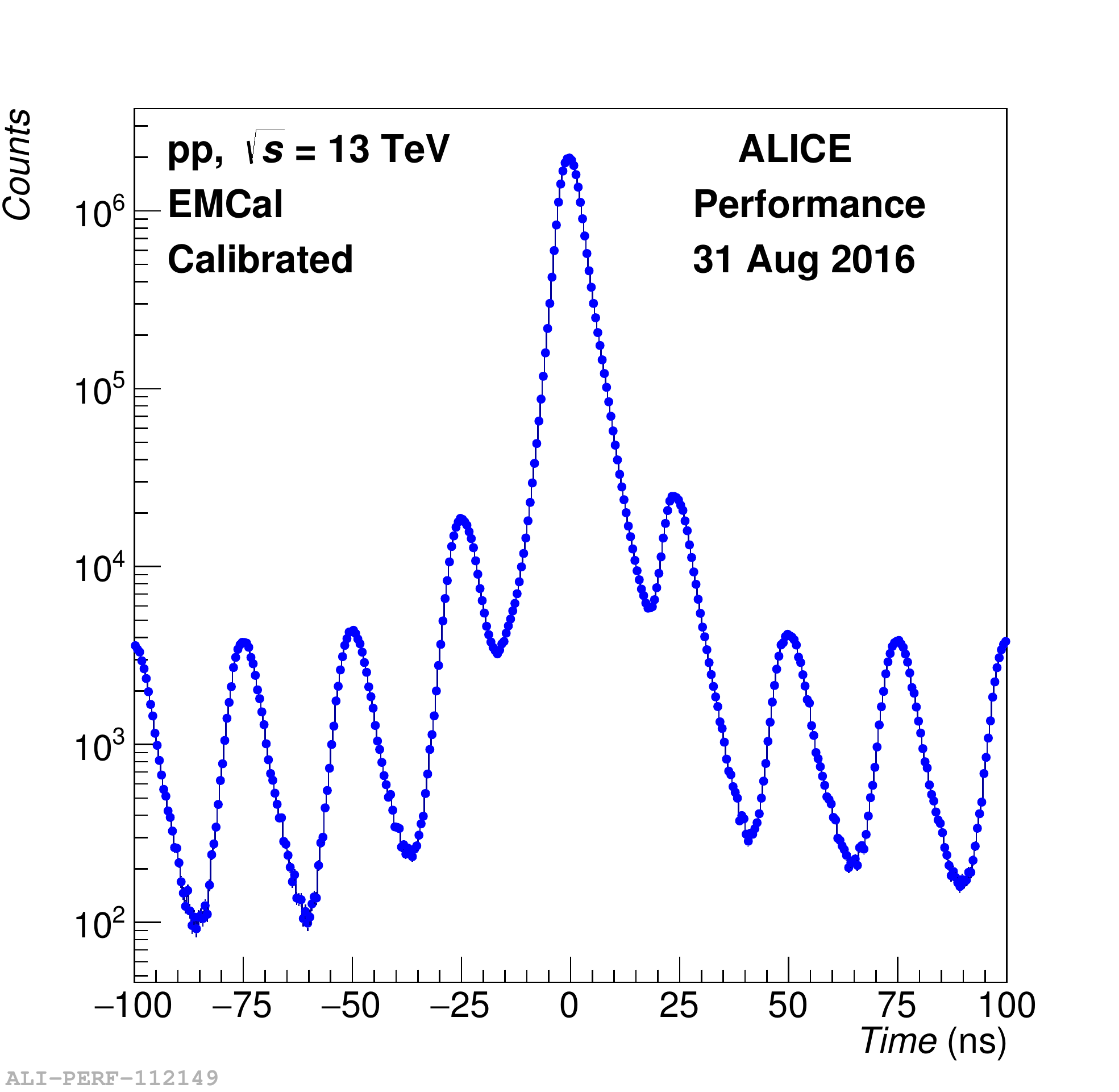}
\caption{Timing distributions in EMCal after timing calibration procedure.}
\label{fig-timing}       
\end{figure}

Minimal bunch crossing (BC) interval at LHC is 25 ns. It is important to have good timing calibration in order to discriminate photons produced in triggered BC and in the surrounding BCs. Front-end electronics (FEE) sampling clock frequency is 10 MHz and is synchronized with LHC clock at 40 MHz, but the phase remains unknown. Both relative cell-by-cell and per-run random phase were calculated from physical data, separately for HG and LG channels. Timing distribution in EMCal after calibration is shown in Figure \ref{fig-timing}, where signals from different BCs are clearly visible.
%
\section{Photon identification}
\label{sec:pid}

\begin{figure}[ht]
\centering
\includegraphics[width=4.9cm]{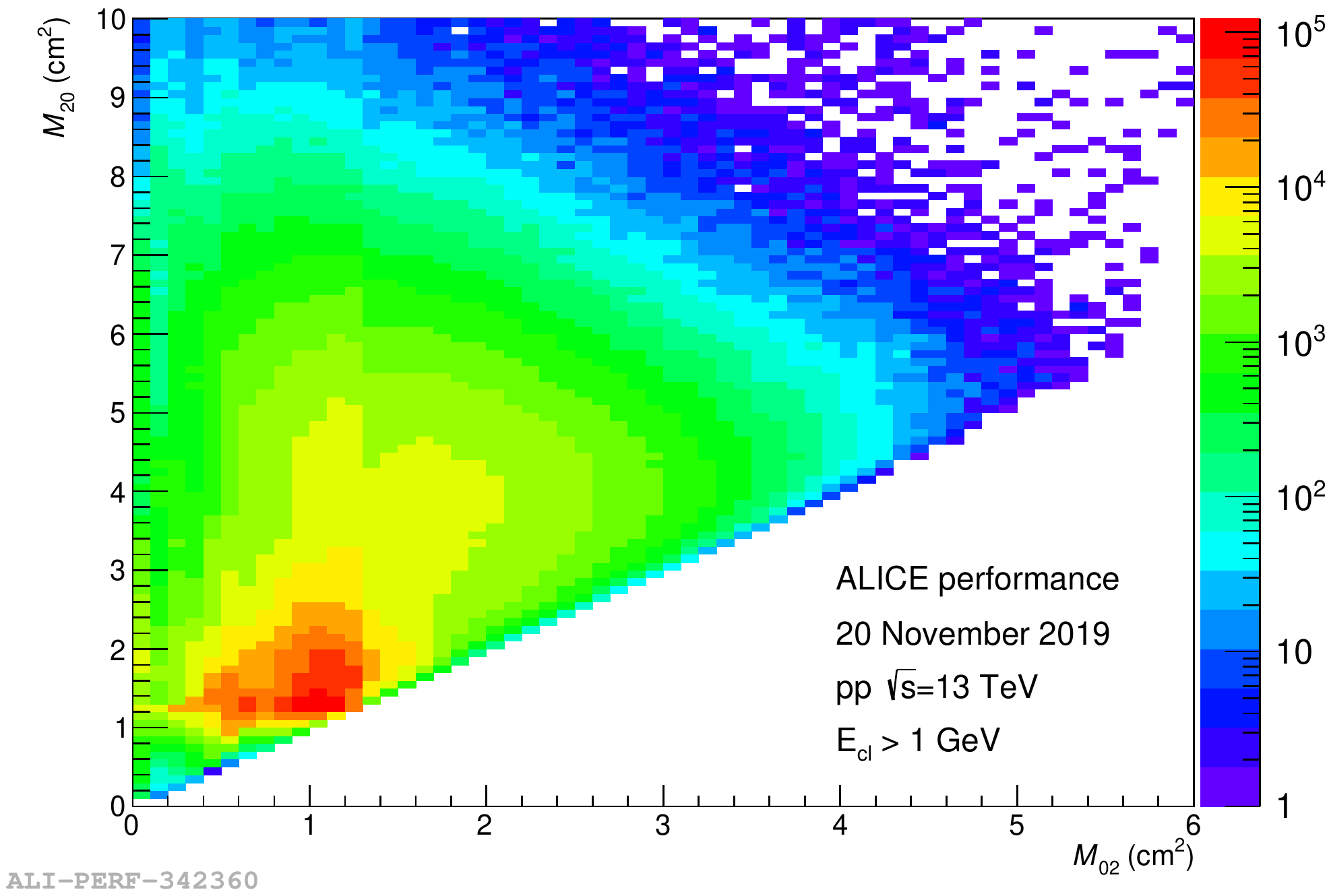}
\includegraphics[width=4.9cm]{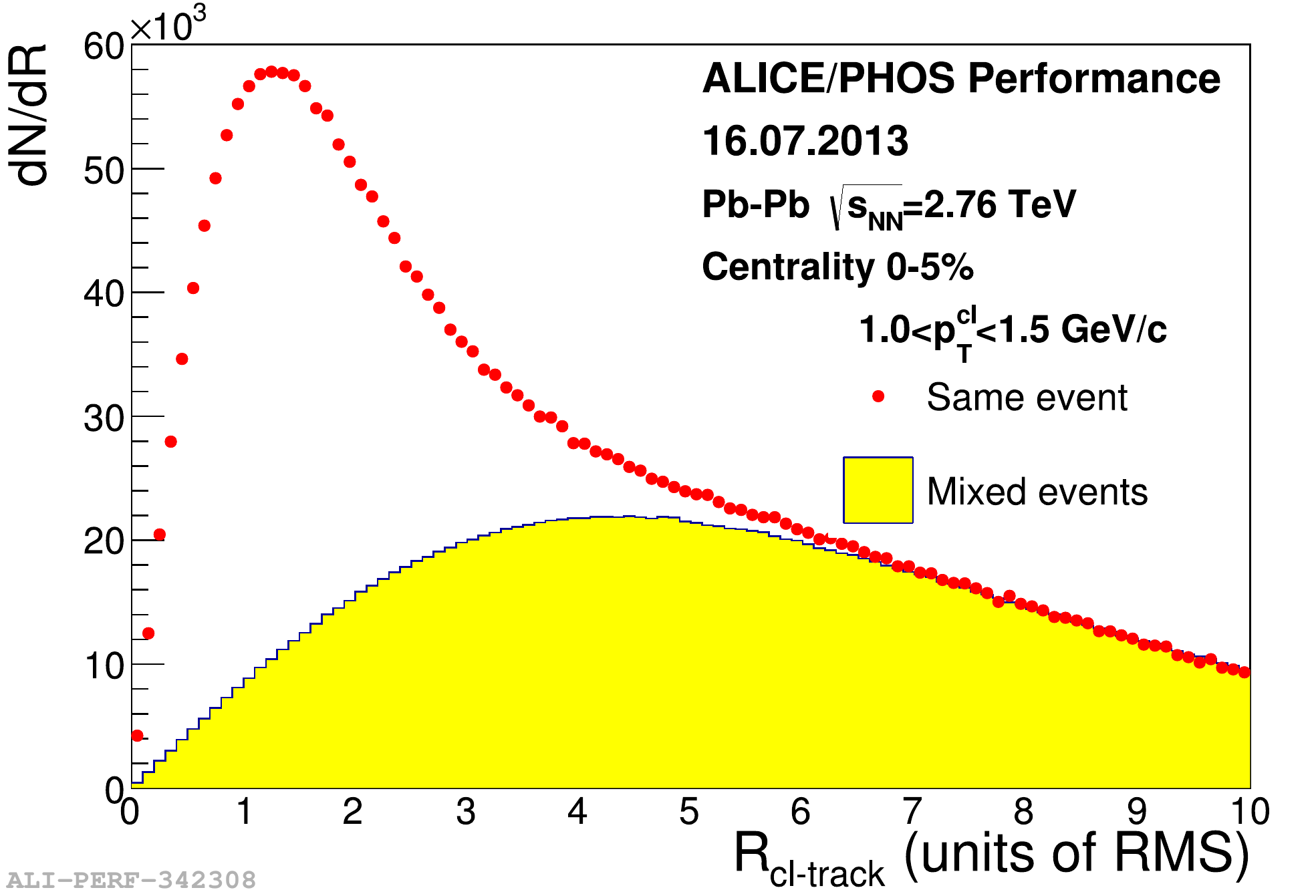}
\includegraphics[width=4.9cm]{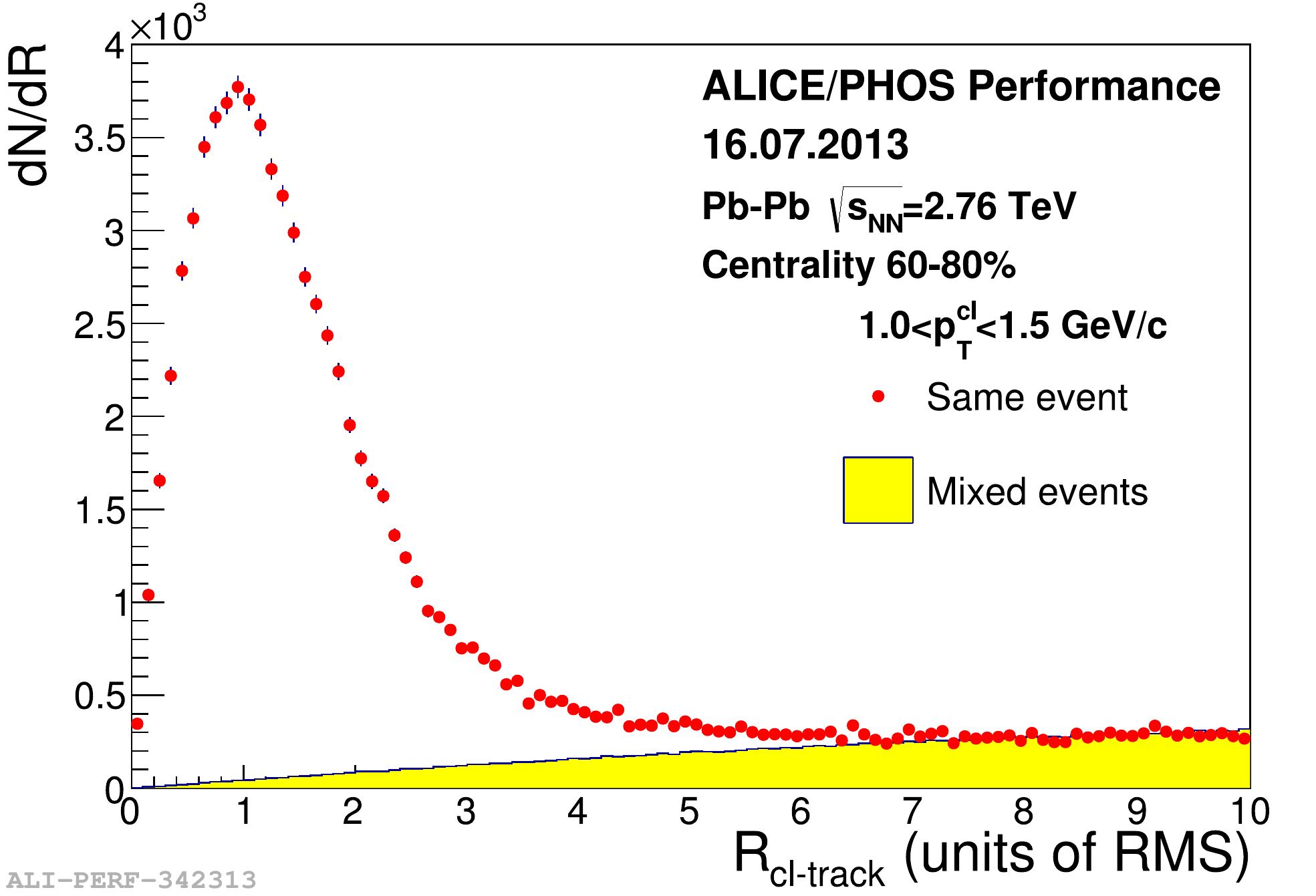}
\caption{Left: distribution of cluster dispersion parameters \cite{ppr2} in PHOS in pp collisions at 13 TeV. Middle and right: distributions of the closest distance between a cluster in PHOS and a track in CTS in two centrality classes of Pb-Pb collisions. Mixed event distributions show combinatorial background. They are obtained by matching tracks and clusters from different events.}
\label{fig-pid}       
\end{figure}

Photon identification in EMCal and PHOS is implemented with two criteria: shower shape and anti-track matching \cite{ppr2}. Shower shape cut is based on 1-dimensional cluster dispersion in EMCal and 2-dimensional one in PHOS. Typical distribution of 2-dimentional dispersion parameters of clusters in PHOS is shown in Figure \ref{fig-pid} (left). Anti-track matching cut is based on distance to the closest extrapolated track from the ALICE central tracking system (CTS) \cite{alice}. Distributions of the closest distance between a cluster in PHOS and a track in CTS in two centrality classes of Pb-Pb collisions are shown in Figure \ref{fig-pid} (middle, right). 
Data provided by CPV detector have been proven to offer the possibilties for improvement of the efficiency of the charged-particle rejection \cite{cpv} and will be fully exploited in the future LHC runs thanks to the enlarged CPV acceptance coverage.
%
\section{Physics performance}
\label{sec:physics}
\begin{figure}[ht]
\centering
\includegraphics[width=5.2cm]{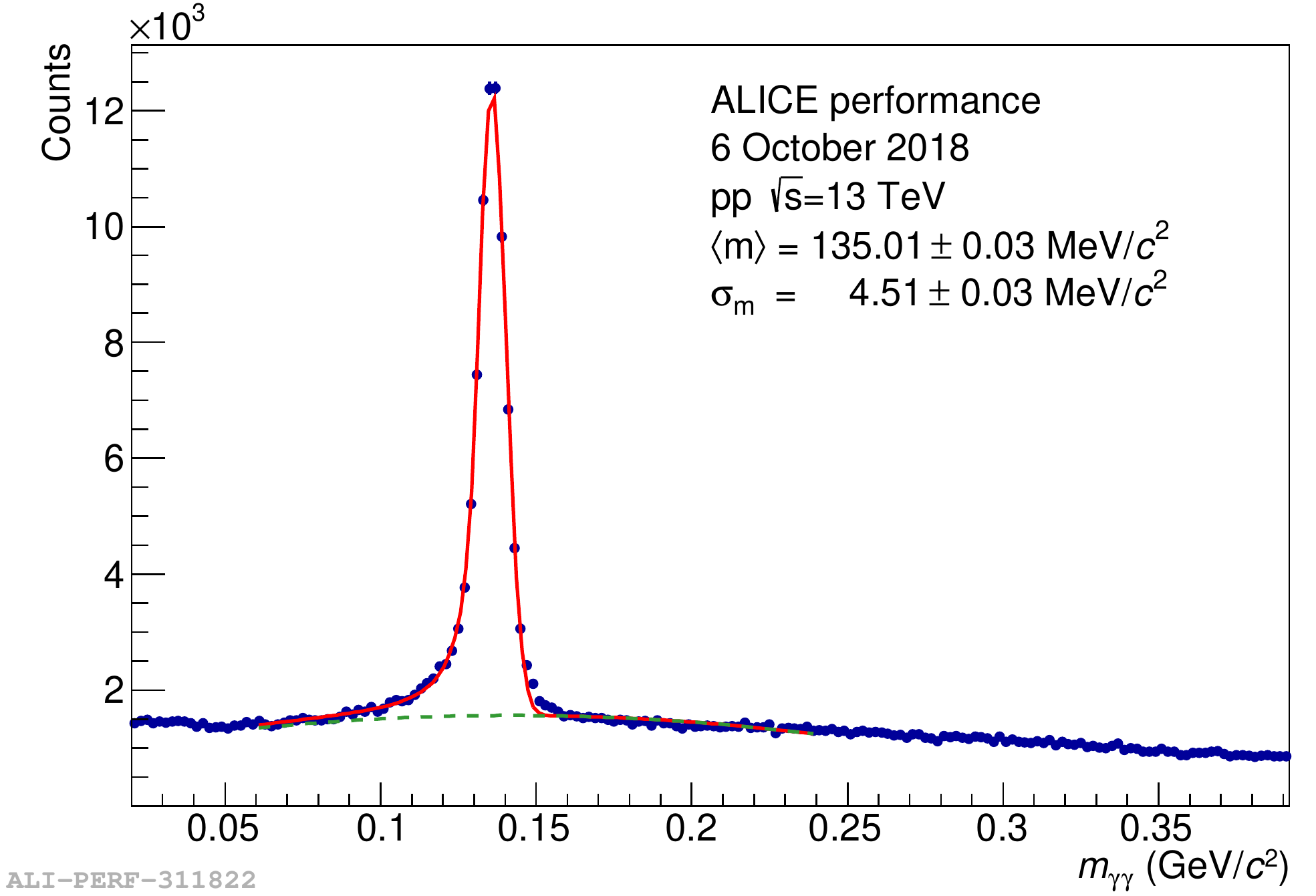}
\includegraphics[width=4.2cm]{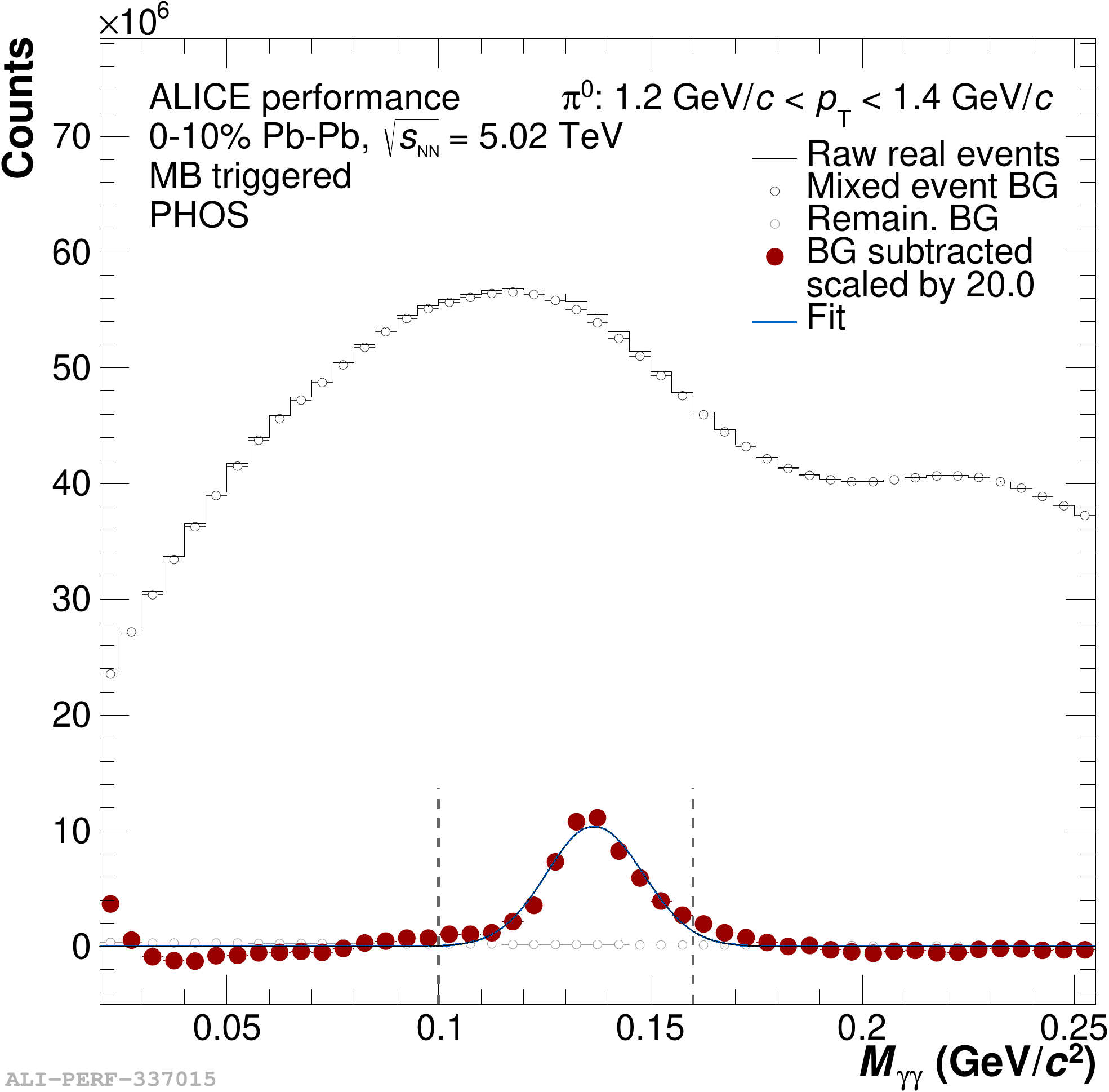}
\includegraphics[width=4.2cm]{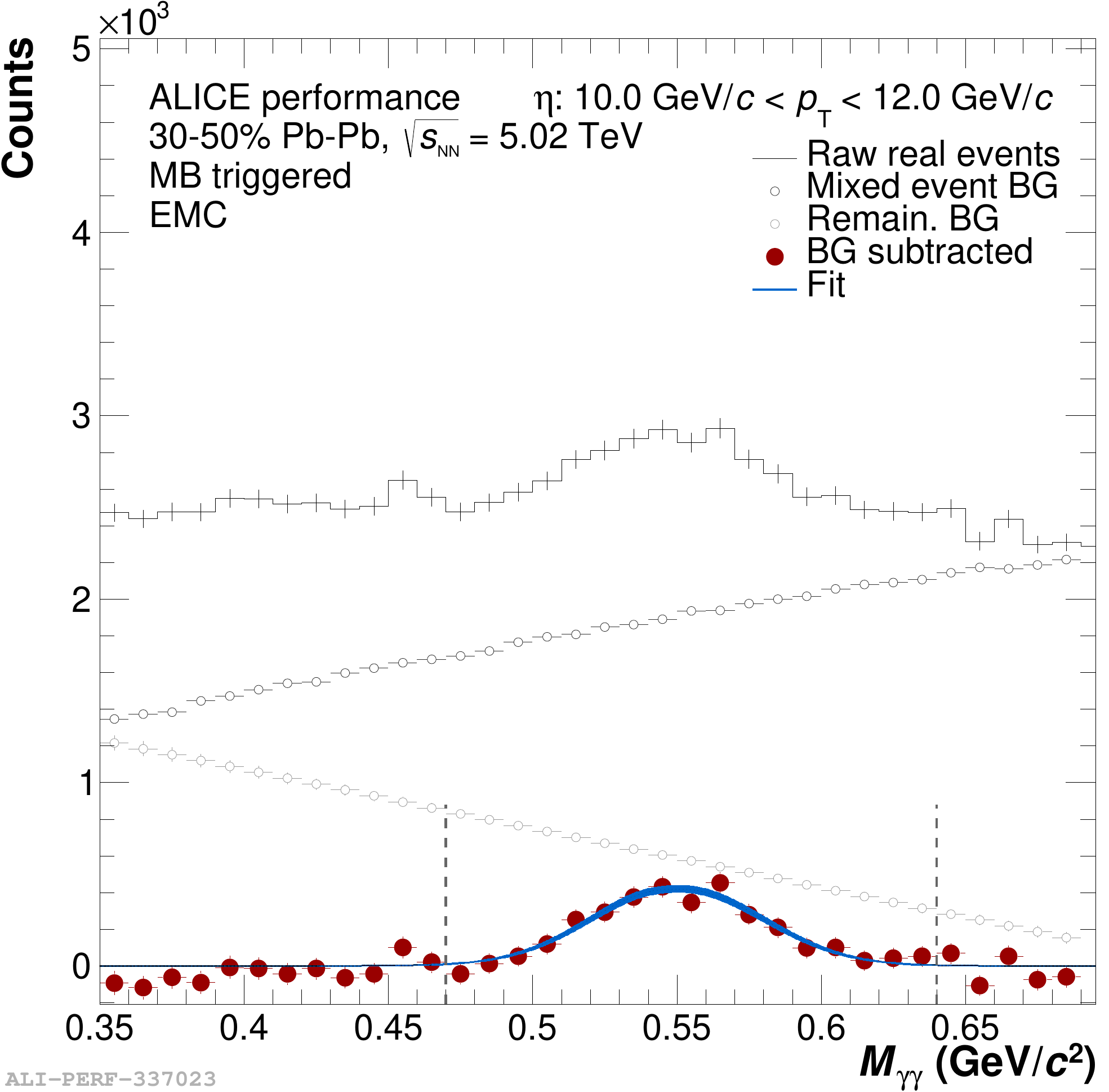}
\caption{Examples of two-photon invariant mass distribution in PHOS and EMCal in pp (left) and Pb-Pb collisions (middle and right).}
\label{fig-pi0}       
\end{figure}

\begin{figure}[ht]
\centering
\includegraphics[width=6.4cm]{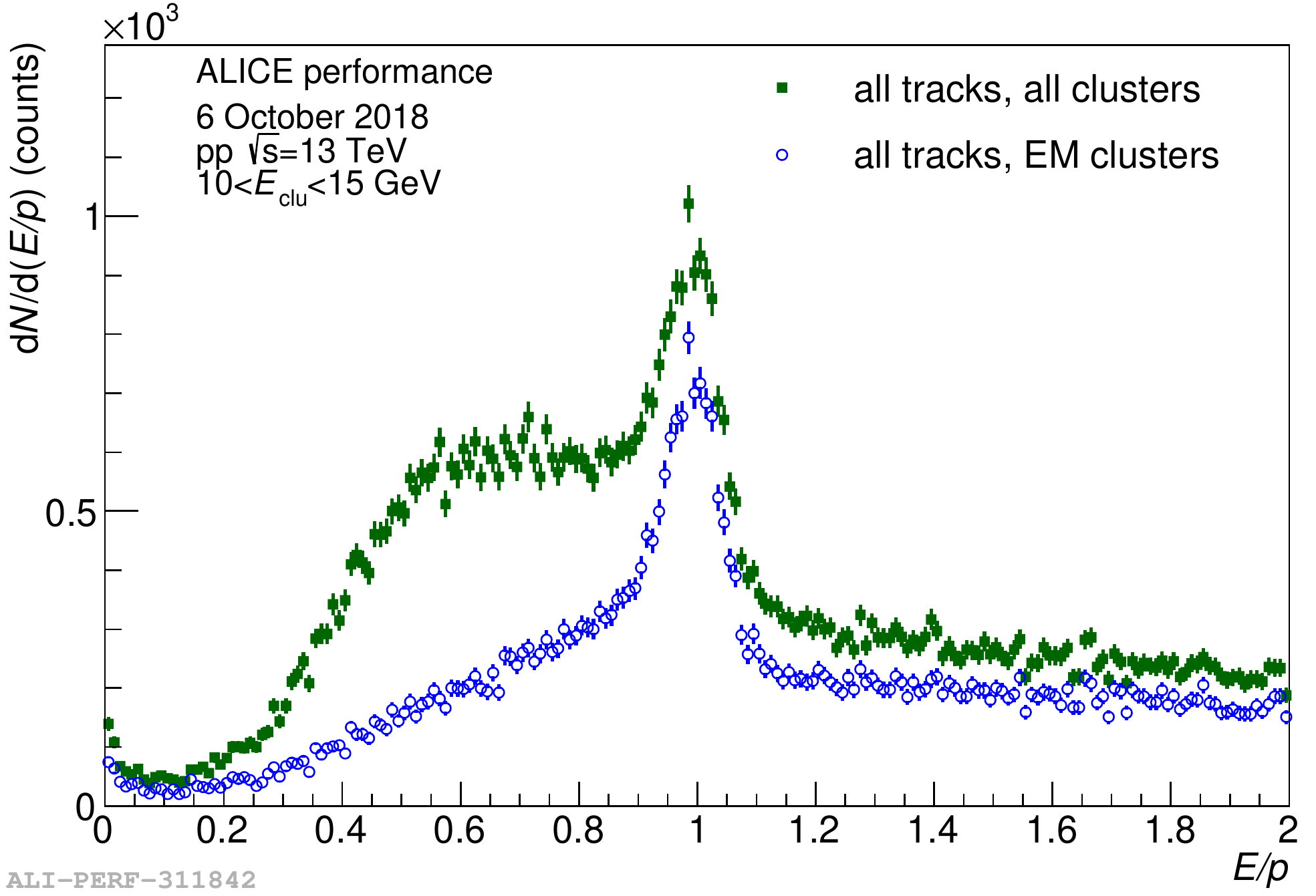}
\includegraphics[width=4.4cm]{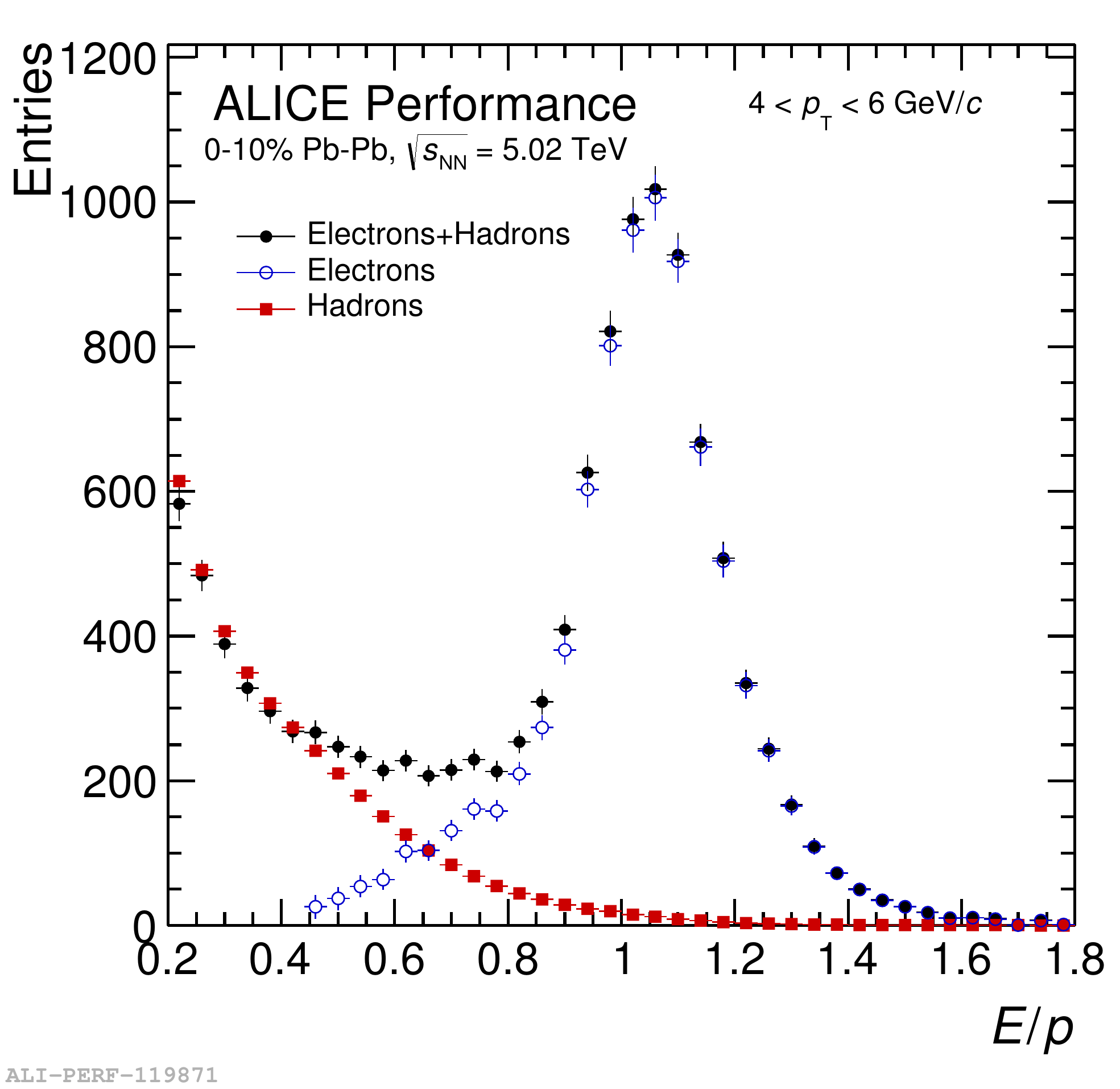}
\caption{The distribution of the cluster energy to track momentum, $E$/$p$ ratio, for PHOS (left) and EMCal (right). A peak around unity due to the electron contribution is visible.}
\label{fig-electron}       
\end{figure}

Precise measurements of $p_\mathrm{T}$ spectra of neutral mesons is important for QCD tests and provide contrains on hadron fragmentation functions and parton distribution functions parameterizations \cite{pi01, pi02}. PHOS and EMCAL are able to reconstruct neutral mesons in low and high multiplicity environments in a wide $p_\mathrm{T}$ ranges. Typical two-photon invariant mass distributions are shown in Figure \ref{fig-pi0}. In pp collisions, excellent resolution of PHOS results in $\pi^0$-peak width as low as $\sigma^{\pi^0}_m = 4.56 \pm 0.03$ MeV/$c^2$ \cite{calib}.
Electrons are identified with high purity in calorimeters by $E$/$p$ ratio in conjunction with tracking system via $dE$/$dx$. Due to wide acceptance and trigger on high-energy electrons, EMCal contributes to electron measurements at high $p_\mathrm{T}$ with high integrated luminosity. PHOS has also the capability to identify electrons. $E$/$p$ distributions for both PHOS and EMCal are shown in Figure \ref{fig-electron}. ALICE results on heavy flavour, i.e. recent measurements of electrons from heavy-flavour hadron decays in p-Pb collisions at $\sqrt{s_{NN}}$ = 5.02 TeV \cite{hf}, heavily rely on EMCal triggering and identification of electrons. One of the goals of EMCal is to trigger on jets and to measure neutral energy of jets. With help of it, fully reconstructed jet spectra are measured in different systems including most recent measurements at 5.02 TeV \cite{jets}.
\section{Upgrade plans}
\label{sec:upgrade}
It is currently anticipated that the acceptance of EMCAL and PHOS will remain unchanched during the future Runs 3 and 4. But in order to comply with high luminosity
requirements of LHC after LS2 upgrade, software for online and offline reconstruction of calorimeters data, data quality control, and calibrations is being prepared within O2 framework \cite{o2}. Firmware in readout and trigger electronics is also being modified in order to reduce the readout time. 

Run 4 upgrade plans for ALICE calorimetry system include an installation of new forward calorimeter FOCAL \cite{focal}. In case of PHOS, R\&D studies are ongoing to upgrade its FEE and photodetectors. Test beam studies \cite{beamtests} showed that the most promising solution will be upgrade from APD of $5\times5$ mm$^2$ size to APD of $10\times10$ mm$^2$ size which will allow to keep current energy resolution parameters without need to cool down the whole detector to -25$^{\circ}$ C. The main advantage of this is a possibility to access the FEE for maintenance during short periods between data taking.

The main purposes of FEE upgrade is to add a timing channel in order to impove time resolution, increase energy dynamic range and produce cards based on modern components. The latter is important for prolongation of the lifespan of the whole detector system. Improvement of time resolution opens possibilities to study direct photons with higher precision. Currently, the prototype of new front-end card is being developed with timing channel on the base of CERN's HPTDC chip \cite{hptdc}. It was tested at CERN during 2017-2018, and energy and time resolutions were measured. Time resolution of about 300 ps was achieved with Hamamatsu multi-pixel photon counter S12572 and about 500 ps with Hamamatu APD S8664-1010 of size $10\times10$ mm$^2$ \cite{beamtests}.

\acknowledgments
This research was supported by the Russian Science Foundation grant 17-72-20234.


\end{document}